\documentclass[aps,preprint,amsmath,amssymb]{revtex4}
\usepackage{graphicx}
\begin{document}

\title{Effect of the excited and fourth generation leptons in lepton flavor violating $\mu^{-} \to e^{-}2\gamma$ decay}

\author{S.C. \.{I}nan}
\email[]{sceminan@cumhuriyet.edu.tr}
\affiliation{Department of Physics, Cumhuriyet University,
58140, Sivas, Turkey}

\begin{abstract}
We study the lepton flavor violating $\mu^- \to e^{-}2\gamma $ decay induced by the mediation of heavy leptons such as the excited and fourth generation leptons. We have examined the effects of these leptons in the $\mu^- \to e^{-}2\gamma $ decay. We have found that branching ratio of this decay is very sensitive to excited and fourth generation leptons. Furthermore, we have obtained the bounds on model parameters by using the current experimental limit of this decay.
\end{abstract}

\pacs{14.80.-j, 1260.Rc}

\maketitle

The Standard Model (SM) has been extremely successful at describing the electroweak interactions. It is in remarkable agreement with the experiments. However, some important questions are still unanswered in SM, such as the proliferation of quarks, leptons generation and their mass spectrum. Two suggested solutions for the these problems are existence of the excited and fourth generation leptons.

SNO collaboration have shown that neutrinos oscillate between families when they propagate over long distances. Therefore, lepton flavor violation can be thought as an experimental fact \cite{sno}. Analysis of the lepton flavor violating $\mu^- \to e^{-}2\gamma $ decay is very interesting due to two different parallel and perpendicular polarization of photons. The examination of the photon spin polarization can provide important information about new physics parameters. With these motivations, we have examined the phenomenology of excited ($\ell_*$) and fourth generation ($\ell_4$) leptons in the lepton violating decay $\mu^- \to e^{-}2\gamma $. In this study, we have assumed that the lepton flavor violation occurs with $\ell_*(\ell_4)-\ell-\gamma$ interaction, in framework of the effective theory. Current experimental limit of this decay is $BR(\mu^- \to e^{-}2\gamma )=7.2\times10^{-11}$ \cite{pdg} and we have used this value when obtaining the bounds on model parameters.

The SM is given by three generations of quark and leptons. A rather natural explanation for these generations is that leptons and quarks composite.  They consist of fewer elementary particles bound by a new strong interaction \cite{baur, boos}. The existence of such quark and leptons substructure should be regarded as the ground state of a rich spectrum of new particles with new quantum numbers. Phenomenologically, an excited lepton can be assumed as a heavy lepton sharing leptonic quantum number with the corresponding SM lepton. Until now have not found any signal for the excited leptons at the  HERA $ep$ \cite{hera1,hera}, LEP $e^{+}e^{-}$ \cite{lep1,lep2,lep3,lep4} and the Tevatron $p\bar{p}$ \cite{tevatron} colliders. However, in these experiments bounds on model parameters have been obtained. Excited leptons have been also studied LHC \cite{lhc, inan} and next $e^{-}e^{+}$, $e\gamma$ colliders \cite{ebo, cak1, cak2, cak3}.

The interaction lagrangian between the SM and excited leptons should conform to the chiral symmetry, therefore light leptons can not acquire radiatively large magnetic moments \cite{bro}. The gauge-mediated effective lagrangian for the anomalous interactions of the excited leptons with SM leptons is \cite{cab, kuhn, hagi, bo1, bo2}

\begin{eqnarray}
L=\frac{1}{2\Lambda}\overline{\ell}^{*}\sigma^{\mu\nu}(gf\frac{\overrightarrow{\tau}}{2}\overrightarrow{W}_{\mu\nu}+g^{'}f^{'}\frac{Y}{2}B_{\mu\nu})\ell_{L}+h.c.
\label{lang1}
\end{eqnarray}
where $\Lambda$ is the compositeness scale which is expected to be $\Lambda<10$ $TeV$ from the experimental and theoretical studies \cite{renard, del, suz}, $g$ and $g^{'}$ are the corresponding electroweak coupling constants, $\overrightarrow{W}_{\mu\nu}$ and $B_{\mu\nu}$ are the field strength tensors of the $SU(2)$ and $U(1)$. $\overrightarrow{\tau}$ and $Y$ are the generators of the corresponding gauge group. The constants $f$ and $f^{'}$ are weight factors associated with the three SM gauge groups which can be defined in a different new physics scale $\Lambda_{i}=\Lambda/f_i$.
In the physical basis, the above Lagrangian can be written in more explicit form

\begin{eqnarray}
\label{lang}
\overline{\ell}^{*}\sigma^{\mu\nu}\ell_{L}+\frac{g_{e}}{2\Lambda}
f\sum_{\ell,\ell^{'}=\nu_{e},e}\Theta^{\overline{\ell}^{*},\ell}_{\mu\nu}\overline{\ell}^{*}\sigma^{\mu\nu}\ell_{L}^{'}+h.c..
\end{eqnarray}
where $N_{\mu\nu}=\partial_{\mu}A_{\nu}-\tan\theta_{W}\partial_{\mu}Z_{\nu}$ is a purely diagonal term
and remaining term is a non-Abelien part which involves triple as well as quartic vertices with

\begin{eqnarray}
\Theta^{\overline{\nu}_e^{*},\nu_{e}}_{\mu\nu}=\frac{2}{\sin2\theta_{W}}\partial_{\mu}Z_{\nu}-i\frac{g_{e}}{\sin^{2}\theta_{W}}W^{+}_{\mu}W^{-}_{\nu},
\end{eqnarray}

\begin{eqnarray}
\Theta^{\overline{e}^{*},e}_{\mu\nu}=-(2\partial_{\mu}A_{\nu}+2\cot2\theta_{W}\partial_{\mu}Z_{\nu}-i\frac{g_{e}}{\sin^{2}\theta_{W}}W^{+}_{\mu}W^{-}_{\nu}),
\end{eqnarray}

\begin{eqnarray}
\Theta^{\overline{\nu}_{e}^{*},e}_{\mu\nu}=\frac{\sqrt{2}}{\sin\theta_{W}}(\partial_{\mu}W^{+}_{\nu}-ig_{e}W^{+}_{\mu}(A_{\nu}+\cot\theta_{W}Z_{\nu})),
\end{eqnarray}

\begin{eqnarray}
\Theta^{\overline{e}^{*},\nu_{e}}_{\mu\nu}=\frac{\sqrt{2}}{\sin\theta_{W}}(\partial_{\mu}W^{-}_{\nu}+ig_{e}W^{-}_{\mu}(A_{\nu}+\cot\theta_{W}Z_{\nu})).
\end{eqnarray}

From equation (\ref{lang}), adding all the contributions, the $V\ell^{*}\ell$  vertex factor is found as follows,

 \begin{eqnarray}
 \label{ver}
 \Gamma^{V\overline{\ell^{*}}\ell}_{\mu}=\frac{g_{e}}{2\Lambda}q^{\nu}\sigma_{\mu\nu}(1-\gamma_{5})f_{V}.
 \end{eqnarray}
Here $V=\gamma, Z, W$ and $q$ is the momentum of the $V$, $f_V$ is the electroweak coupling parameter and defined for photon as $f_{\gamma}=Q_{f}f^{'}+I_{3L}(f-f^{'})$. In this study we have assumed $f=f^{'}$.

\begin{figure}
\includegraphics{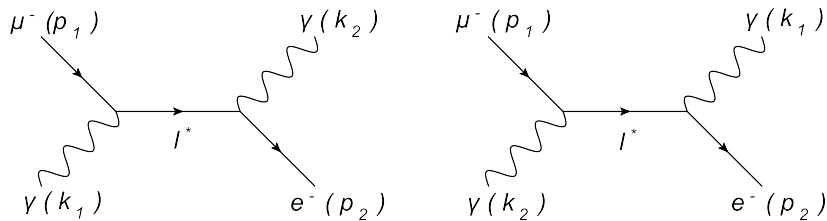}
\caption{Feynman Diagrams for $\mu^{-}\to e^{-}2\gamma$  . \label{fig1}}
\end{figure}

The Feynman diagrams for the $\mu^{-} \to e^{-}2\gamma $ are shown in Fig.\ref{fig1}. The matrix element squared for this process can be obtained as follows,

\begin{eqnarray}
|M|^2=2g_e^{4}\left(\frac{f_\gamma}{\Lambda}\right)^4m_*^2s^2\beta(m_{\mu}^2+s(1-\beta z)) \nonumber \\
\times\left(\frac{1}{(q_1^2-m_*^2)^2}+\frac{1}{(q_2^2-m_*^2)^2}+\frac{2}{(q_1^2-m_*^2)(q_2^2-m_*^2)}\right)
\label{amp}
\end{eqnarray}
This calculation have been performed with using the center of mass frame of the outgoing photons. Here $s=(p_1-p_2)^2=(k_1+k_2)^2$, $q_1^2=(p_1-k_1)^2=(p_2+k_2)^2$, $q_2^2=(p_1-k_2)^2=(p_2+k_1)^2$, $\beta=m_\mu^2/s-1$, $z=cos(\theta)$ and $m_*$, $m_\mu$ are masses of the excited lepton and muon respectively. We have neglected the mass of the electron. The partial decay widths $d\Gamma$ can be obtained with using Eq.\ref{amp} as

\begin{eqnarray}
\frac{d\Gamma}{dsdz}=\frac{1}{2^{9}\pi^3m_\mu^2}\ \left(\frac{s}{E_\mu}\right)\ \beta\ \frac{1}{2}\ \frac{1}{2} |M|^2,
\end{eqnarray}
where $E_{\mu}=(s+m_\mu^2)/(s\sqrt{s})$ is the muon energy in corresponding mass frame.

\begin{figure}
\includegraphics{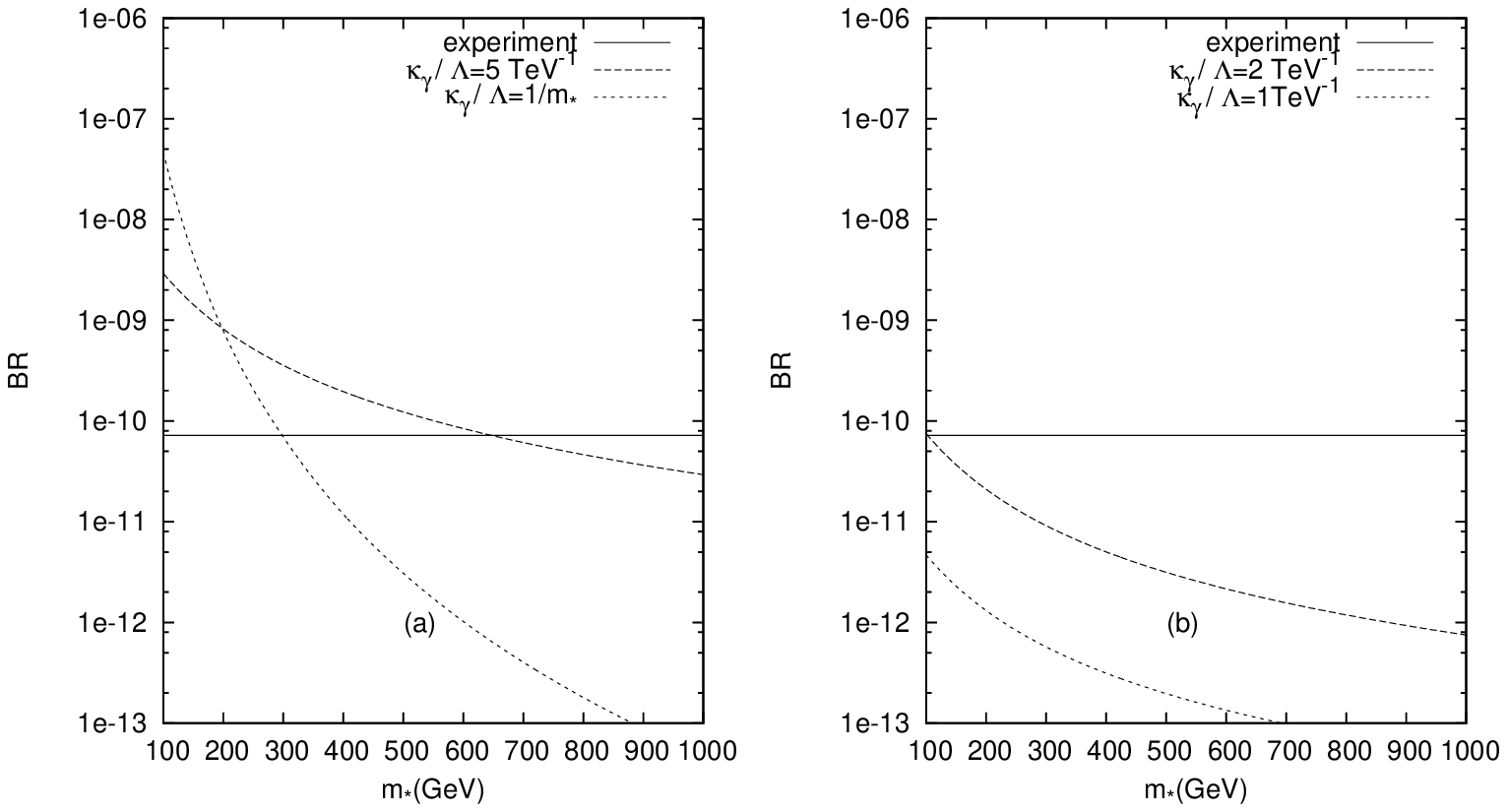}
\caption{Variation of the branching ratio (BR) with $m_*$ for (a) $f_{\gamma}/\Lambda=5$ $TeV^{-1}$, $1/m_*$ and (b) $f_{\gamma}/\Lambda=1,2$ $TeV^{-1}$   .\label{fig2}}
\end{figure}

\begin{figure}
\includegraphics{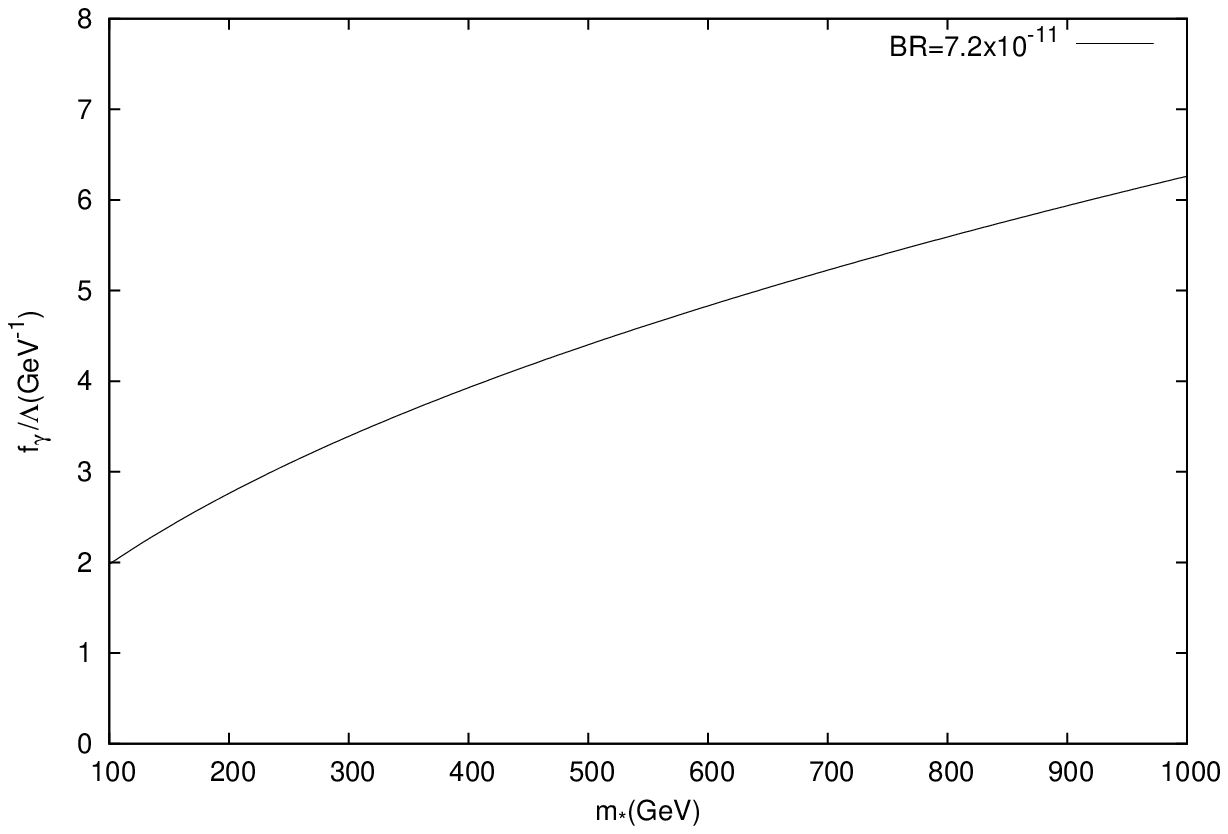}
\caption{ Exclusion area for the $m_*$ and $f_{\gamma}/\Lambda$ when the $BR(\mu \to e_-2\gamma)=7.2\times10^{-11}$. The excluded area is over the curve.\label{fig3}}
\end{figure}

In Fig.\ref{fig2}a,b the magnitude of the $m_*$ for different $f_{\gamma}/\Lambda$ values is shown. As we can see from the figure, the branching ratio is very sensitive to the $m_*$ and $f_{\gamma}/\Lambda$. We have obtained lower bounds on the $m_*>650$ GeV for $f_{\gamma}/\Lambda=5$ $TeV^{-1}$, $m_*>302$ GeV for  $f_{\gamma}/\Lambda=1/m_*$ and $m_*>102$ GeV for $f_{\gamma}/\Lambda=2$ $TeV^{-1}$ as can be seen from the figure. Fig.\ref{fig3} represents the excluded area for the $m_*$ and $f_{\gamma}/\Lambda$ when the $BR(\mu \to e^{-}2\gamma)=7.2\times10^{-11}$. The excluded area is over the curve.

The flavor democracy hypothesis suggests the fourth generation family in SM \cite{dat, cel, sul}. Due to large mass of the $t$ quark, its anomalous interactions should be enhanced compared to that of the other SM fermions as discussed in Ref.\cite{hf}. Therefore, fourth generation leptons $(\ell_4)$ masses are expected to be heavy \cite{atag1} and therefore it can be expected that can make a remarkable contribution to new physics. As discussed in \cite{riz, cif}, the effective lagrangian for the anomalous interactions of the fourth family charged leptons with photon can be written as follows,

\begin{eqnarray}
L=\left(\frac{\kappa_{\gamma}^{\ell_{i}}}{\Lambda}\right)e_{l}g_{e}\overline{\ell_{4}}\sigma_{\mu\nu}\ell_{i}F^{\mu\nu}+ h.c.
\end{eqnarray}
where $\kappa_{\gamma}^{\ell_{i}}$ is the anomalous coupling for the neutral currents with a photon, $\Lambda$ is a new physics energy scale, $e_{l}$ is the electric charge of the leptons, $g_e$ is the electromagnetic coupling
constant, $F^{\mu\nu}$ is field stress tensor of the photon and $\sigma_{\mu\nu}=i[\gamma_{\mu},\gamma_{\nu}]/2$.

\begin{figure}
\includegraphics{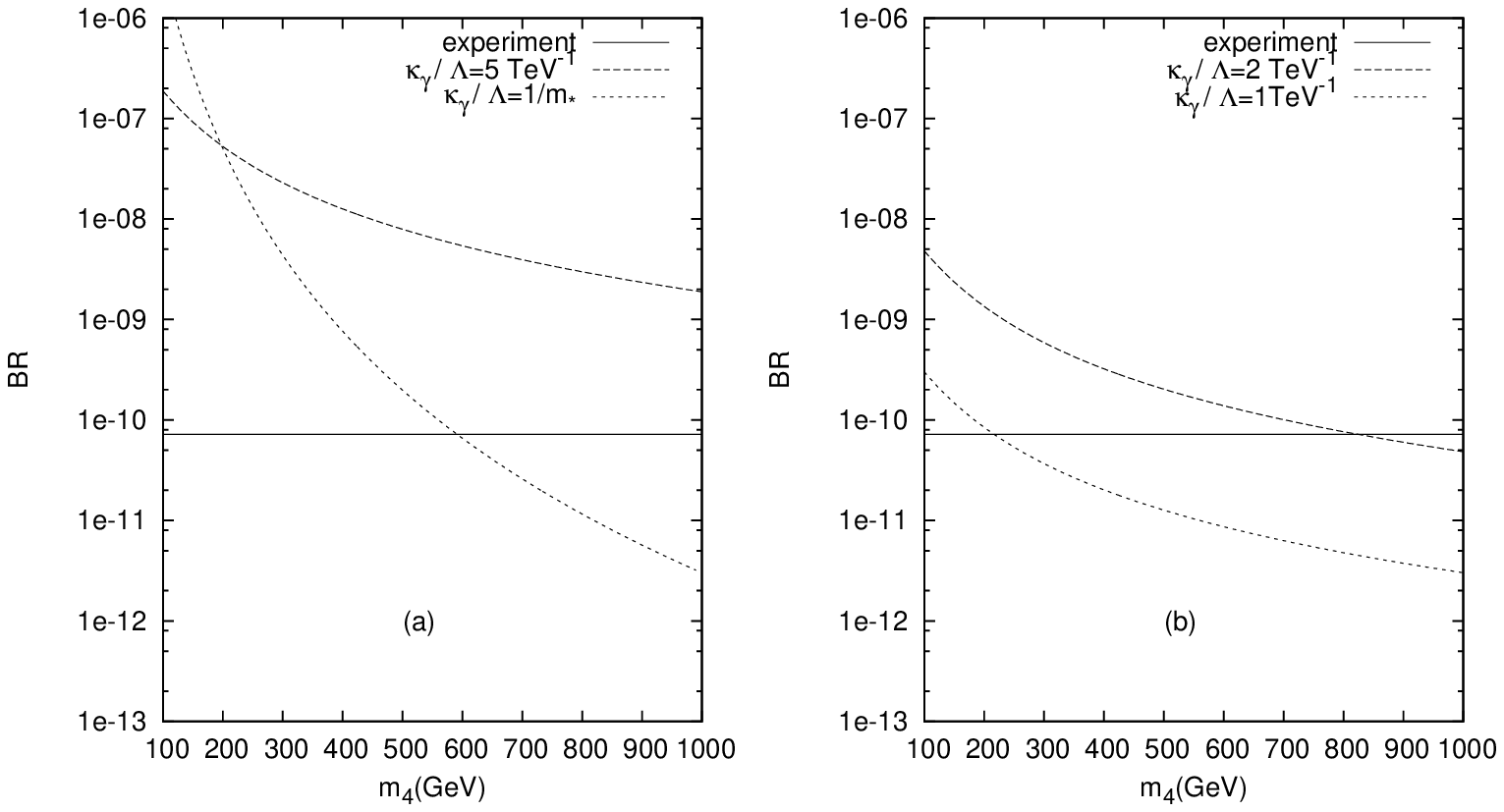}
\caption{Variation of the branching ratio (BR) with $m_4$ for (a) $\kappa_{\gamma}/\Lambda=5$ $TeV^{-1}$, $1/m_*$ and (b) $f_{\gamma}/\Lambda=1,2$ $TeV^{-1}$.
\label{fig4}}
\end{figure}

\begin{figure}
\includegraphics{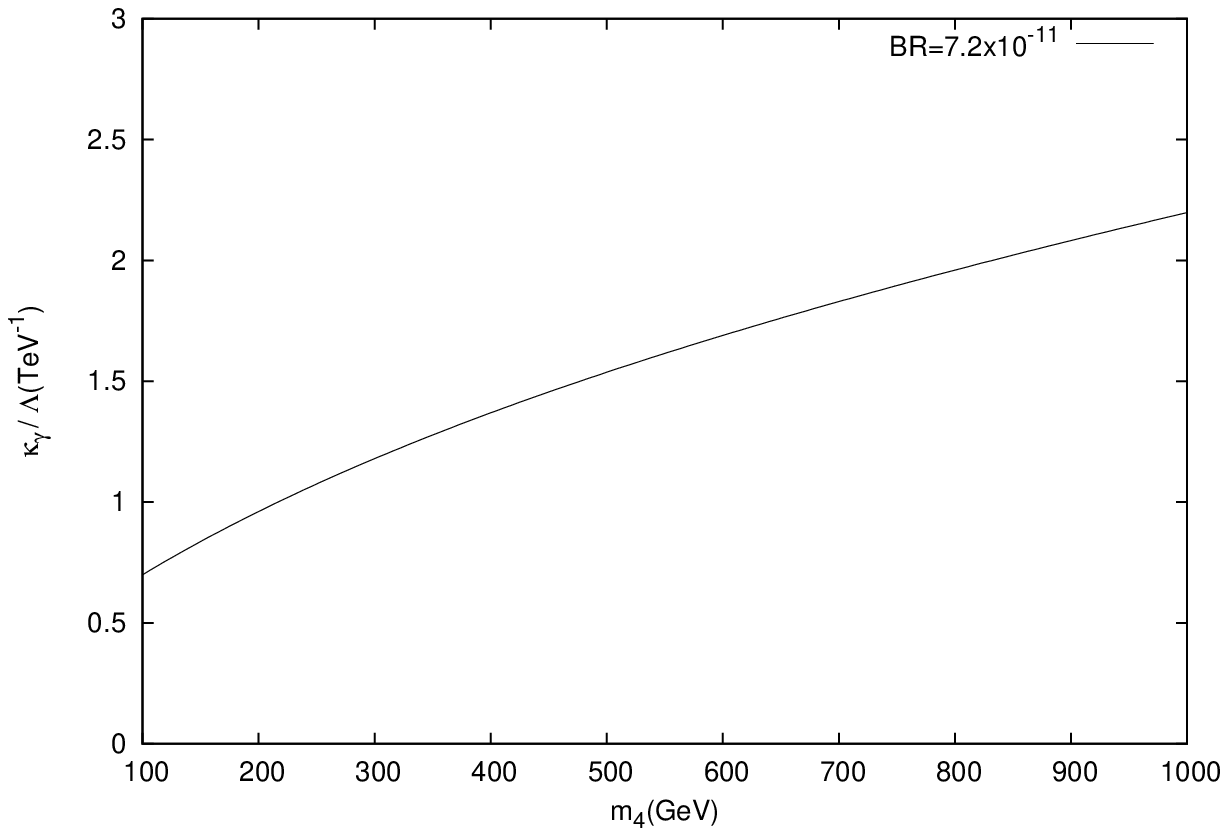}
\caption{ Exclusion area for the $m_4$ and $\kappa_{\gamma}/\Lambda$ when the $BR(\mu \to e_-2\gamma)=7.2\times10^{-11}$. The excluded area is over the curve.\label{fig5}}
\end{figure}

The Feynman diagrams for the $\mu^{-} \to e^{-}2\gamma $  induced by fourth generation leptons can be obtained by replacements $\ell_* \to \ell_4$ in Fig.\ref{fig1}. The whole polarization summed amplitude has been calculated in two photons mass frame as,

\begin{eqnarray}
|M^2|=16g_e^{4}\left(\frac{\kappa_\gamma}{\Lambda}\right)^4 s^2\beta(1+z)(2m_4^2+m_\mu^2-s(1-\beta z))(m_\mu^2+s(1-\beta z)) \nonumber \\ \times\left(\frac{1}{(q_1^2-m_4^2)^2}+\frac{1}{(q_2^2-m_4^2)^2}+\frac{2}{(q_1^2-m_4^2)(q_2^2-m_4^2)}\right)
\end{eqnarray}
 where $m_4$ is the mass of the fourth generation lepton and other physical parameters are same as those discussed in section II. In Fig.\ref{fig4}a, we present the magnitude of the $m_4$ for different $\kappa_{\gamma}/\Lambda$. Similar to excited leptons, the branching ratio is very sensitive to the new physics parameters; $m_4$ and $\kappa_{\gamma}/\Lambda$. We have obtained lower bounds on the $m_4>2000$ GeV (not shown in the figure) for $\kappa_{\gamma}/\Lambda=5$ $TeV^{-1}$, $m_4>590$ GeV for  $\kappa_{\gamma}/\Lambda=1/m_4$ from the Fig.\ref{fig4}a and $m_4>805$ GeV for $\kappa_{\gamma}/\Lambda=2$ $TeV^{-1}$, $m_4>198$ GeV for $\kappa_{\gamma}/\Lambda=1$ $TeV^{-1}$ from the Fig.\ref{fig4}b. Fig.\ref{fig5} represents the excluded area for the $m_4$ and $\kappa_{\gamma}/\Lambda$ when the $BR(\mu^{-} \to e^{-}2\gamma)=7.2\times10^{-11}$. The excluded area is over the curve.

The lepton flavor violating decay $\mu^{-} \to e^{-}2\gamma$ can  allow us to understand the new physics idea. In addition, this decay can exist in tree level with the existence of excited and fourth generation leptons.

Initially, we have introduced the effect of the excited leptons on $\mu^{-} \to e^{-}2\gamma$ decay. As can be seen from the Fig.\ref{fig2}, branching ratio of $\mu^{-} \to e^{-}2\gamma$ strongly depends on the excited leptons. Obtained bounds on model parameters of the excited lepton complementary to current experimental bounds \cite{hera1, hera, lep1, lep2, lep3, lep4, tevatron}. In table \ref{tab1}, we have recapitulated the limits at HERA, LEP and Tevatron in comparison with our result for  $f_{\gamma}/\Lambda=1/m_*$. The expected LHC limits ($480-1000$ GeV) are much stronger than our limits \cite{lhc, inan}. However, with the more accurate future experiments, it would be possible to check the effects of excited leptons and provide us important information about the bounds on parameters of this model.

Next, we have examined the fourth generation leptons. The branching ratio of $\mu^{-} \to e^{-}2\gamma$ decay is shown in Fig.\ref{fig4}. It can be seen from this figure that branching ratio is very sensitivity to fourth generation leptons. There is no experimental results for the fourth family leptons yet. However, the bounds on $m_4$ have been obtained next $ep$ and $e\gamma$ colliders \cite{cif, cif2}. In these studies, obtained bounds on $m_4$ lies between $500-1500$ GeV for different $\kappa_{\gamma}/\Lambda$ and collider energies. Our obtained limits are comparable with these limits.

As a result, measurements of upper limits of branching ratio's of $\mu^{-} \to e^{-}2\gamma$ can give important information for the excited and fourth generation leptons.

\pagebreak

\pagebreak

\begin{table}
\caption{The obtained limits at HERA, LEP and Tevatron with their references and obtained limits in this paper. Lower bounds of $m_*$
are given in units of GeV.
\label{tab1}}
\begin{ruledtabular}
\begin{tabular}{cccc}
$f_\gamma/\Lambda=1/m_*$ \\
HERA &LEP &Tevatron &BR($\mu^{-} \to e^{-}2\gamma$) \\
\hline
$228-272$           &$103-295$                       &$221$        &$302$ \\
\cite{hera1,hera}   &\cite{lep1,lep2,lep3,lep4}  &\cite{tevatron}  \\
\end{tabular}
\end{ruledtabular}
\end{table}

\end{document}